\documentclass[useAMS,usenatbib,usegraphicx,usehyperref]{mn2e}

\newcommand{\beq}{\begin{equation}}
\newcommand{\eeq}{\end{equation}}
\newcommand{\beqa}{\begin{eqnarray}}
\newcommand{\eeqa}{\end{eqnarray}}

\title[Magnetic inhibition of macro-turbulence]{On magnetic inhibition
  of photospheric macro-turbulence generated in the iron-bump opacity
  zone of O-stars}

\author[Sundqvist et al.]{J.O. Sundqvist$^{1,2}$\thanks{E-mail:
    mail@jonsundqvist.com}, V. Petit$^{2}$, S.P. Owocki$^{2}$,
  G.A. Wade$^3$, J. Puls$^1$,\newauthor and the MiMeS Collaboration\\ $^1$Institut f\"ur
  Astronomie und Astrophysik der Universit\"at M\"unchen,
  Scheinerstr. 1, 81679 M\"unchen, Germany\\ $^2$University of Delaware, Bartol
  Research Institute, Newark, Delaware 19716, USA\\ $^3$Royal Military
  College of Canada, Department of Physics, PO Box 17000 Kingston,
  Ontario K7K 7B4, Canada\\}

\begin{document}

\date{Accepted 2013-05-22. Received 2013-04-29}

\pagerange{\pageref{firstpage}--\pageref{lastpage}} \pubyear{2002}

\maketitle

\label{firstpage}

\begin{abstract}
  Massive, hot OB-stars show clear evidence of strong macroscopic
  broadening (in addition to rotation) in their photospheric spectral
  lines. This paper examines the occurrence of such
  ``macro-turbulence'' in slowly rotating O-stars with strong,
  organised surface magnetic fields. Focusing on the C\,{\sc iv}
  5811\,\AA \ line, we find evidence for significant macro-turbulent
  broadening in all stars except NGC\,1624-2, which also has (by far)
  the strongest magnetic field. Instead, the very sharp C\,{\sc iv}
  lines in NGC\,1624-2 are dominated by magnetic Zeeman broadening,
  from which we estimate a dipolar field $\sim$\,20\,kG. By contrast,
  magnetic broadening is negligible in the other stars (due to their
  weaker field strengths, on the order of 1\,kG), and their C\,{\sc
    iv} profiles are typically very broad and similar to corresponding
  lines observed in non-magnetic O-stars. Quantifying this by an
  isotropic, Gaussian macro-turbulence, we derive $v_{\rm mac} = \,
  \rm 2.2 \pm^{0.9}_{2.2}\, km/s$ for NGC-1624, and $v_{\rm mac}
  \approx 20-65 \, \rm km/s$ for the rest of the magnetic sample.

  We use these observational results to test the hypothesis that the field 
  can stabilise the atmosphere and suppress the generation of
  macro-turbulence down to stellar layers where the magnetic pressure
  $P_{\rm B}$ and the gas pressure $P_{\rm g}$ are comparable. Using a
  simple grey atmosphere to estimate the temperature $T_0$ at which
  $P_{\rm B} = P_{\rm g}$, we find that $T_0 > T_{\rm eff}$ for all
  investigated magnetic stars, but that $T_0$ reaches the
  $\sim$\,160\,000\,K layers associated with the iron opacity-bump in hot
  stars only for NGC\,1624-2. This is consistent with the view that
  the responsible physical mechanism for photospheric O-star
  macro-turbulence may be stellar gravity-mode oscillations excited by
  sub-surface convection zones, and suggests that a sufficiently
  strong magnetic field can suppress such iron-bump generated
  convection and associated pulsational excitation.

\end{abstract}


\section{Introduction}
\label{intro}


Modern high-resolution, high signal-to-noise optical spectroscopy of
hot stars show clearly that rotation is not the only macroscopic
line-broadening agent operating in their photospheres. The additional
broadening is typically seen over the complete spectral line, has a
Gaussian-like shape and very large widths, $\sim$\,50\,km/s, well in
excess of the photospheric speed of sound, $\sim$\,20\,km/s. Indeed,
such very large ``macro-turbulence''\footnote{In stellar spectroscopy,
  one typically distinguishes between such \textit{macro}-turbulence,
  with scales longer than the photon mean-free-path (mfp), and
  \textit{micro}-turbulence, with scales shorter than this mfp. For
  O-stars, inferred micro-turbulent velocities are normally on order a
  few km/s, i.e. much smaller than the macro-turbulence that is the
  focus of this paper.} seems to be a ubiquitous feature of OB-star
atmospheres \citep{Howarth97, SimonDiaz07, Lefever07, SimonDiaz10b,
  Markova11, Najarro11, Bouret12}.

However, since early-type stars lack the vigorous surface convection
associated with hydrogen recombination -- which is responsible for
such additional, non-thermal broadening in late-type stellar
atmospheres \citep{Asplund00} -- the physical origin of
macro-turbulence in hot stars remains unclear. A long-standing
suggestion has been stellar oscillations \citep{Lucy76}, and recently
\citet{Aerts09} showed that the collective line-broadening effect of
numerous low-amplitude gravity-mode (g-mode) pulsations can indeed
mimic the large inferred values of Gaussian-like macro-turbulence.
Stellar structure theory predicts that this pulsational broadening
excited by \textit{core convection} becomes significant only in
evolved massive stars, and while first results for OB supergiants do
indeed seem to support a pulsational origin of macro-turbulence
\citep{SimonDiaz10b}, observations indicate the presence of
significant macro-turbulent velocities in hot stars of \textit{all}
evolutionary classes (\citealt{Markova11}; \citealt{Najarro11};
Sim{\'o}n-D{\'{\i}}az et al. in prep; Markova et al., in
prep.). Recently, however, \citet{Shiode13} \citep[see
  also][]{Cantiello09} suggested that \textit{near-surface} convection
generated by the iron-bump opacity zone can trigger excited g-modes
in main-sequence O-stars as well, and that such pulsations might be the
origin of their observed turbulent broadening.

A key empirical issue regards the actual quantification of
macro-turbulence. In particular, additional broadening due to a
basically asymmetric process like stellar oscillations will affect the
first zero in the Fourier power spectra of spectral lines, which typically 
is used to infer the projected stellar rotation \citep[e.g.,][]{SimonDiaz07}, 
and so make it difficult to disentangle and quantify the relative amounts
of macro-turbulent and rotational broadening present in the atmosphere
\citep{Aerts09}. Here we circumvent this issue by focusing on O-stars
with detected organised surface magnetic fields, using high-quality
spectra collected within the Magnetism in Massive Stars project
\citep[MiMeS,][]{Wade12}. In particular, all these magnetic O-stars
have very long measured rotation periods, likely because they have been 
spun down through magnetic braking by their strong stellar winds
\citep[see][]{Petit13}; their (in most cases) negligible rotational
broadening makes them ideal targets for the present study.

From studies of the chemically peculiar Ap stars, it is well known
that the presence of a sufficiently strong surface magnetic field can
stabilise the atmosphere against large-scale motions and suppress
convectively induced turbulent line-broadening (J. Landstreet,
T. Ryabchikova, O. Kochukhov, priv. comm., see also, e.g.,
\citet{Michaud70}; \citealt{Landstreet96}). Building on these
concepts, we compare here the narrow lines observed in NGC\,1624-2
(measured surface dipolar field $\sim$\,20\,kG, \citealt{Wade12b})
with the broader lines observed in other magnetic O-stars (surface
fields on order $\sim$\,1\,kG), in an attempt to place first
observational constraints on the depth of the source region for the
enigmatic macro-turbulence in hot main-sequence stars.

The paper is organised as follows: sect.~\ref{Observations} presents
the observational sample and describes our method for deriving
macro-turbulent velocities. Sect.~\ref{results} gives the results of
the analysis, and Sect.~\ref{inhibition} presents a simple model for
interpreting magnetic inhibition of macro-turbulence. Finally,
Sect.~\ref{discussion} discusses the results, gives our conclusions,
and outlines directions for future work.

\section{Observations and method}
\label{Observations}

We select magnetic O-type stars with well constrained rotational
periods and strong magnetic fields from the compilation by
\citet{Petit13}. The sample, listed in Table \ref{Tab:params},
consists of 7 magnetic stars, including NGC\,1624-2 which hosts an
order of magnitude stronger field than any other known magnetic O star
\citep{Wade12b}. For comparison purposes, we also include and analyse
the non-magnetic star HD\,36861 (=$\lambda$\,Ori\,A), which has
stellar parameters similar to those of our magnetic sample\footnote{Spectropolarimetric 
  observations available for this star constrain the dipolar
  surface field to be below 80\,G (David-Uraz et al. in
  prep).}.

For all sample stars, we retrieve high-quality spectra from the
extensive MiMeS database. The observations were obtained with the
high-resolution ($R\sim65\,000$) spectropolarimeter ESPaDOnS at the
Canada-France Hawaii Telescope, or with its twin instrument Narval
located at the T\'elescope Bernard-Lyot. We focus on analysing the
magnetic O-stars' low states, defined according to when the cyclic
H$\alpha$ emission originating from their ``dynamical magnetospheres''
\citep{Sundqvist12b} is at minimum. This should minimise any
contamination of the photospheric lines by magnetospheric emission
\citep[for a review of massive-star magnetospheric properties,
  see][]{Petit13}. However, at the end of Sect.~\ref{results} we also
comment briefly on line-profile variability between this low state and
the corresponding high state (defined by maximum H$\alpha$ emission).

To obtain macro-turbulent velocities we analyse the photospheric
C\,{\sc iv} 5811\,\AA \ line, accounting also for rotational and
magnetic broadening when needed. Surface magnetic field properties of
OB stars are generally derived from polarimetric measurements, as the
Zeeman splitting typically is too small to be separated from other
broadening mechanisms \citep[see][]{Donati09}. In the optical, the
Zeeman broadening is only $\sim$\,1-2\,km/s per kG, i.e. in the
current sample only NGC\,1624-2 has a surface field strong enough to
display significant magnetic broadening in the line profiles (see
Figure~\ref{Fig:fits}). Nevertheless, for consistency we include a
proper treatment of the Zeeman effect for all stars, using the
literature dipolar field values listed in Table \ref{Tab:params}. We
note that, due to the sometimes unknown observer inclination angle
$i$, these inferred field strengths are uncertain by up to $\sim
50$\,\%; this uncertainty, however, does not affect the
macro-turbulent velocities derived in this paper. The Zeeman pattern
of the C\,\textsc{iv} $3s\,^2S_{1/2} \rightarrow 3p\,^2P_{1/2}$
transition, calculated under $LS$ coupling, has two strong $\sigma$
components and two well-spaced, weaker $\pi$ components. This pattern,
along with a high magnetic sensitivity (effective Land{\'e} factor
$\bar{g}=1.33$), makes this unblended line particularly suitable for
the present analysis.

The rotation period of magnetic stars can be readily obtained from the
observed variation of the longitudinal field \citep[e.g.,][]{Borra80,
  Bychkov05} or photometric/spectral variations caused by their
magnetospheres \citep[e.g.,][]{Landstreet78, Howarth07}. Most of the
magnetic O-stars have very long periods and so have negligible
rotational broadening, except for HD\,148937 and HD\,37022
(=$\theta^1$ Ori C), for which we include broadening due to the modest
rotation listed in Table \ref{Tab:params}.

We compute emergent intensities at each point on the stellar surface
using the Unno-Rachkovsky solution for a Milne-Eddington atmosphere
(\citealt{Unno56}; see also Ch. 9 in \citealt{Landi04}). Synthetic
flux profiles are then obtained by numerically integrating the
emergent intensities over the projected stellar disc, assuming a
dipolar surface field matching the simple geometry of the (presumed
fossil) fields observed in these stars. Formally, the Zeeman splitting
is sensitive to the surface field modulus; however, for equal
disk-integrated values of the modulus the actual position of the
magnetic pole on the visible hemisphere does not change the structure
of the line profile significantly. Therefore we place the magnetic 
equator at the center of the stellar disc for all stars, as 
the best representative viewing angle for their low states.  

This simplified approach allows for a fast treatment of the Zeeman
effect and derivation of macro-turbulent velocities, and further
avoids model dependencies on e.g. stellar parameters and chemical
abundances, which are not important for the analysis here. Within the
Unno-Rachkovsky approximation, the source function $S$ is linear in
the continuum optical depth,
$S(\tau_\mathrm{c})=S_0[1+\beta\tau_\mathrm{c}$], and the polarised
radiative transfer equations can be solved analytically. Specifically,
we use $\beta=1.5$, Voigt-shaped line profiles with damping constant
$a=0.02$, and a thermal speed $v_{\rm th} = 10$\,km/s. We then fit the
core of the line over a grid of profile strengths (set by the
line-to-continuum opacity ratio) and isotropic\footnote{In some studies \citep[e.g.,][]{SimonDiaz10b,
    Najarro11} a radial-tangential formulation of macro-turbulence is
  used rather than the isotropic one adopted here. In the former case,
  large-scale motions are assumed to occur only radially and/or
  tangentially to the stellar surface \citep[see][]{Gray05}.}, Gaussian
macro-turbulence of the form $e^{-v^2/v_{\rm mac}^2}/(\sqrt{\pi}
v_{\rm mac})$. To
estimate the admissible $v_{\rm mac}$, we calculate the Bayesian
probability density function marginalised over the
profile-strengths. The prior probability density is constant, and any
deviations not explained by the model are treated as additional
Gaussian noise, which results in the most conservative estimate of the
parameters under the maximum entropy principle
\citep[e.g.,][]{Gregory05}.


\section{results} 
\label{results}

\begin{table*}
\begin{minipage}{\textwidth}
    \centering
    \caption{Literature stellar and magnetic parameters, and median
      macro-turbulent velocities derived in this paper, including
      errors derived from the 68\,\% confidence regions.}
        \begin{tabular}{ l l l l l l l l l l }
        \hline \hline Star             &Obs. date&Inst. & Spec. type &$T_{\rm eff}$ & $\log g$     & $B_{\rm pole}$ &$P_\mathrm{rot}$ & $v \sin i$ & $v_{\rm mac}$ \\
                                 &&&& [\,kK\,] & [\,cgs\,]                 & [\,kG\,] &[d]& [\,km\,s$^{-1}$\,] & [\,km\,s$^{-1}$\,] \\ \hline
                $^a$NGC\,1624-2         &2012-09-27& ESP & O6.5-O8\,f?cp & 35 & 4.0 & 20     &158& 0 & $2.2\ \ \pm\,^{0.9}_{2.2}$ \\ 
                $^b$HD\,191612         &2008-08-19& ESP & O6\,f?p-O8\,f?cp & 35 & 3.5 & 2.5     &538& 0 & $62.0\,\pm\,^{0.5}_{0.5}$ \\
                 $^c$HD\,57682             & 2010-01-29 & ESP & O9\,V & 34 & 4.0 & 1.7     &64& 0 & $19.2\,\pm\,^{0.3}_{0.3}$ \\
                $^d$CPD\,-28\,2561     &2012-02-10& ESP & O6.5\,f?p & 35 & 4.0 & 1.7     &70& 0 & $24.3\,\pm\,^{1.0}_{0.9}$ \\
                $^e$HD\,37022         &2006-01-08& ESP & O7\,Vp & 39 & 4.1 & 1.1     &15& 24 & $42.9\,\pm\,^{0.5}_{0.6}$ \\
                $^f$HD\,148937         &2010-07-23& ESP & O6\,f?p & 41 & 4.0 & 1.0     &7& 45 & $54.0\,\pm\,^{0.9}_{0.9}$ \\
                $^g$HD\,108             &2009-07/2009-08& Nar& O8\,f?p & 35 & 3.5 & 0.5     &1.8\,$\times 10^4$& 0 & $64.4\,\pm\,^{0.4}_{0.4}$ \\
                $^h$HD\,36861         &2007-21-21& ESP & O8\,III((f)) & 35 & 3.7 & 0     &--& 45 & $50.0\,\pm\,^{0.3}_{0.3}$ \\
                \hline
                \multicolumn{10}{l}{$^a$\citet{Wade12b} $^b$\citet{Wade11a}  $^c$\citet{Grunhut12} $^d$Barba et al. in prep; \citet{Petit13}}  \\    
                \multicolumn{10}{l}{$^e$\citet{Wade06} $^f$\citet{Wade12c} $^g$\citet{Martins10}  $^h$\citet{Najarro11}} \\    
        \end{tabular}
    \label{Tab:params}
\end{minipage}
\end{table*}

\begin{figure}
  \includegraphics[width=0.48\textwidth]{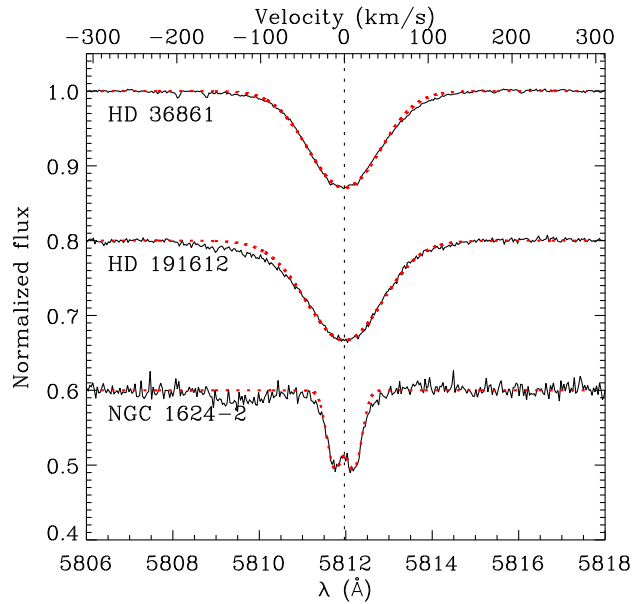}
  \caption{Observed (black solid) and fitted (red dashed) {\sc c\,iv}
    line profiles for three stars in our sample, as labeled in the
    figure. The horizontal dashed line marks line center, and the
    continua in the two lower curves have been shifted downwards by 0.2
    and 0.4 normalized flux units.}
  \label{Fig:fits}
\end{figure}

\begin{figure}
  \includegraphics[width=0.48\textwidth]{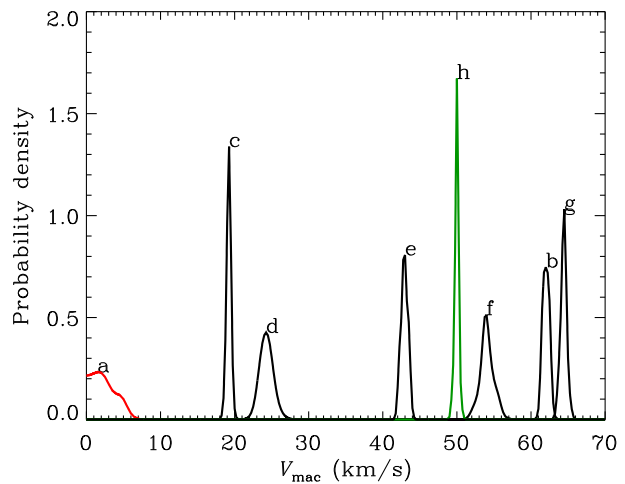}
  \caption{Marginalised probability density distributions of derived
    macro-turbulent velocities, labelled according to Table 1. The red
    curve indicates NGC\,1624-2, and the green curve indicates the
    non-magnetic comparison star HD\,36861.}
  \label{Fig:vmac}
\end{figure}

\begin{figure}
  \includegraphics[width=0.48\textwidth]{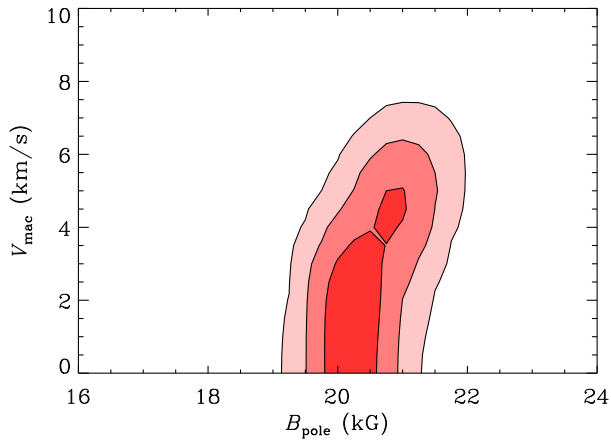}
  \caption{Joint probability-density contour map of derived
    macro-turbulent velocity and magnetic dipolar field strength for
    NGC\,1624-2. The contours encircle 68.3, 95.4, and 99.7 per cent
    of the probability.}
  \label{Fig:contour}
\end{figure}

Using the method described above, Figure~\ref{Fig:fits} shows fitted
{\sc c\,iv} profiles for three stars in our sample, namely HD\,191612,
NGC\,1624-2, and the non-magnetic comparison star HD\,36861.

Since HD\,191612's rotation period is 538 days and its surface
magnetic field is ``only'' 2.5\,kG (Table 1), both rotational and
magnetic broadening are negligible for this star; thus the total line
broadening may be quite unambiguously associated purely with
macro-turbulence. Comparing this with HD\,36861 reveals very similar
Gaussian-like and broad C\,{\sc iv} lines, indicating a common origin
of the observed macro-turbulence in magnetic and non-magnetic
O-stars. By contrast, the observed line in NGC\,1624-2 is
qualitatively very different, much narrower and with magnetic Zeeman
splitting directly visible (due to the very strong surface
field). This indicates that the mechanism responsible for the large
macro-turbulent velocities in HD\,36861 and HD\,191612 is not
effective in NGC\,1624-2.

Figure~\ref{Fig:vmac} further shows probability-density distributions
of macro-turbulent velocities for the full stellar sample described in
Sect.~\ref{Observations}, and Table 1 lists the corresponding median
values along with the 68\,\% confidence regions. For NGC\,1624-2, we
allow the (dipole) magnetic field strength to be a free parameter, in
order to properly model the observed Zeeman splitting and obtain a
more conservative estimate of the macro-turbulent
velocity. Figure~\ref{Fig:contour} illustrates the resulting
joint-probability contour-map of magnetic field strength and
macro-turbulence.

This simple analysis for NGC\,1624-2 results in a very strong dipole
field $B_{\rm pole} \approx 20$\,kG (consistent with
\citealt{Wade12b})
and a very low $v_{\rm mac} = \, \rm 2.2 \pm^{0.9}_{2.2}\,km/s$. Such
a low macro-turbulent velocity is in stark contrast with the rest of
the sample, which displays much higher values, $v_{\rm mac} \approx \,
\rm 20-65\, km/s$ (see Figure~\ref{Fig:vmac}). Thus, among the known
slowly rotating magnetic O-stars, macro-turbulence is anomalously 
low in NGC\,1624-2.

We have here focused on fitting profiles during the spectral ``low
state'' (see Sect.~\ref{Observations}). Comparing the observed C\,{\sc
  iv} lines in the low and high states of the magnetic stars indeed
reveals variable and asymmetric profiles, but it is difficult to judge
whether this variability is of photospheric origin or stems from
refilling by magnetospheric emission during the high state. Thus we
defer a detailed investigation of such line-profile variability to a
future paper, noting simply that the inferred macro-turbulent
velocities during the low and high states typically differ only by a
few km/s, which does not affect our basic results or conclusions.


\section{Interpreting magnetic inhibition of O-star macro-turbulence}
\label{inhibition}

Let us now interpret the results of the previous section using a very
simple model for magnetic inhibition of macro-turbulence. Assuming the
magnetic field stabilises the atmosphere against large-scale motions
approximately down to a stellar layer at which the magnetic pressure
$P_{\rm B} = B^2/(8 \pi)$ equals the gas pressure $P_{\rm g}$, we can
analytically estimate the corresponding temperature $T_0$ by adopting
a classical grey atmosphere with temperature structure at given
optical depth
\begin{equation}
  T(\tau) \approx T_{\rm eff} \left( \frac{3}{4} \tau + \frac{1}{2}  
  \right)^{1/4}.
  \label{Eq:T0}
\end{equation} 
For gravity $g$ (in $\rm cm/s^2$) and a nearly constant mass
absorption coefficient $\kappa$ (in $\rm cm^2/g$), the optical depth
becomes
\begin{equation}
  \tau = \int \kappa \rho dz \approx \kappa m_{\rm c} = \kappa P_{\rm
    g}/g,
\end{equation} 
where the last equality uses hydrostatic equilibrium to evaluate the
column mass $m_{\rm c}$. For a large-scale magnetic field with
arbitrary, and possibly even variable, tilt to the vertical direction,
the variation of density along a field line depends only on the
vertical depth. With the further assumption that there is no
horizontal variation in pressure or density between field lines, this
standard hydrostatic condition between pressure and column mass still
holds, i.e. $P_{\rm g} = m_{\rm c} g$.

Since the extent of the photosphere is very small compared to the
stellar radius, we may further neglect the radial dependence of the
magnetic field and use the inferred surface field strength $B$
throughout the atmosphere\footnote{Note also here that since these
  large-scale, organised magnetic fields presumably are of fossil
  origin, they should not exhibit the increase in field strength with
  density that may be expected for fields in energy equipartition
  generated by stellar dynamos.}. Setting then $P_{\rm g} = P_{\rm B}
= B^2/(8 \pi)$, one obtains the temperature $T_0$ in the layer at
which the gas and magnetic pressure are equal,

\begin{equation}
  T_0 = T_{\rm eff} \left( \frac{3}{32 \pi} \frac{B^2 \kappa}{g} + 
  \frac{1}{2} \right)^{\frac{1}{4}}
  \approx 0.42 \, T_{\rm eff} B^{\frac{1}{2}} (\kappa/g)^{\frac{1}{4}},
  \label{Eq:T0}
\end{equation} 
where $B$ has units of Gauss. The second expression here neglects the
1/2 within the parenthesis, and so implicitly assumes a field strength
significantly stronger than the $B \approx 400 \, (10^{-4}
g/\kappa)^{1/2}$ that yields $P_{\rm B} = P_{\rm g}$ at $T_0 = T_{\rm
  eff}$.

Table 1 lists stellar and magnetic parameters used to estimate $T_0$ for 
the stars in our sample. We use the averaged surface field for $B$ 
and, for simplicity, we further take $\kappa = 1 \approx 3 \,
\kappa_{\rm e}$ for all stars, where $\kappa_{\rm e} = 0.34$ is the
electron scattering opacity for a hot star of standard solar
composition. Inspections of Rosseland opacities in detailed {\sc
 fastwind} non-LTE model atmospheres \citep{Puls05} show that for
atmospheric layers with $\tau_{\rm Ross} \ga 0.1$, such constant
$\kappa \approx 3\,\kappa_{\rm e}$ actually is a quite good
opacity-estimate in Galactic O-stars that are not too evolved (see
also, e.g., Fig. 1 of \citealt{Cantiello09}).

\begin{figure}
  \includegraphics[width=0.48\textwidth]{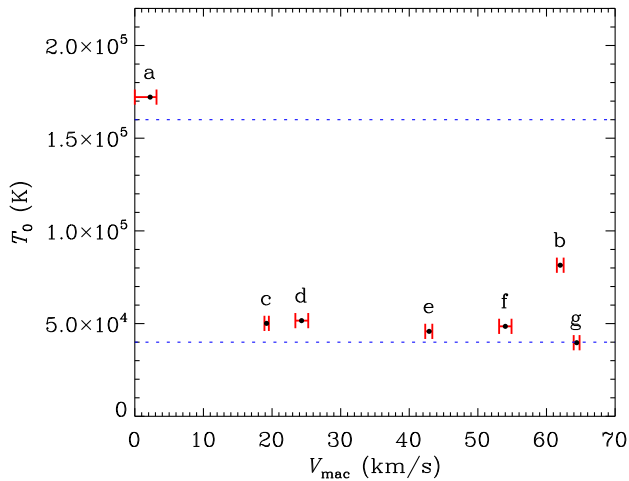}
  \caption{Atmospheric temperature $T_0$ (defined as the temperature
    at which gas and magnetic pressure equals according to the model
    described in text) vs. derived macro-turbulent velocities with
    1$\sigma$ error bars. Stars labeled according to Table 1. The
    dashed lines denote the approximate locations of the stellar
    effective temperatures (lower) and the iron opacity-bump (upper).}
  \label{Fig:T0}
\end{figure}

Figure~\ref{Fig:T0} displays $T_0$ vs. the $v_{\rm mac}$ values
derived in Sect. 3 for the magnetic O stars. The figure shows that
although the magnetic field likely affects the atmosphere of all
investigated stars to some extent, this influence reaches down to much
deeper layers in the sharp-lined NGC\,1624-2 than in any other star.
A simple comparison of NGC\,1624-2 ($v_{\rm mac} =
2.2\,\pm\,^{0.9}_{2.2}\rm \,km/s$) with HD\,191612 ($v_{\rm mac} =
62.0\,\pm^{0.5}_{0.5}\rm \,km/s$), which gives the second highest
$T_0$, then suggests that the physical mechanism causing the large
values of macro-turbulence likely originates in stellar layers with $
100\,000 \rm \, K \la \it T \la \rm 200\,000 \, K$. This is consistent
with a physical origin of O-star photospheric macro-turbulence in the
iron-peak opacity zone located roughly at $T \approx 160\,000$\,K.

\section{discussion and conclusions}
\label{discussion}

The central result of this paper is that strong macro-turbulence is
present in the photospheres of all known slowly rotating magnetic O
stars except for NGC\,1624-2, which has (by far) the strongest surface
field, and in which such non-thermal line-broadening seems to be
largely suppressed. Moreover, the broad lines observed in the other
magnetic O stars are typically quite similar to those seen in
non-magnetic stars, suggesting a common origin of macro-turbulence. As
in non-magnetic stars \citep[e.g.,][]{Markova11}, it is not clear why
these less extreme magnetic stars show such a large range of
macro-turbulent velocities.

Using these unique constraints derived from the magnetic O stars, we
assume (in analogy with Ap stars) that the organised, presumed fossil
field can inhibit large-scale atmospheric motions down to stellar 
layers where the magnetic and gas pressures are comparable. The
analysis here then indicates that the physical mechanism responsible
for macro-turbulence in O stars likely originates in atmospheric layers 
with temperatures corresponding to $100\,000\,\rm K \la \it T \rm \la 200\,000\,K$.

An attractive scenario then is that this physical mechanism may be
stellar oscillations excited by convection in the near-surface,
iron-bump opacity zone of hot stars \citep{Shiode13}. Future
theoretical studies should calculate the excitation of such
pulsational modes and their effects on spectral line formation,
extending the work by \citet{Aerts09} who computed line-profiles from
the collective effect of more deep-seated g-mode pulsations in evolved
hot stars.

Finally we note also that the iron opacity-bump becomes weaker for
lower metallicity \citep{Cantiello09}. Thus, if indeed sub-surface
convection is ultimately responsible for the observed non-thermal line
broadening in early-type main-sequence stars, it would imply that
lower metallicity stars should have lower macro-turbulence. This
should be observationally examined by comparing inferred
macro-turbulent velocities in the Galaxy and the Magellanic Clouds.

\section*{Acknowledgments}

We thank Sergio Sim{\'o}n-D{\'{\i}}az, Matteo Cantiello, John
Landstreet, Tanya Ryabchikova, and Oleg Kochukhov for valuable
discussions. JOS gratefully acknowledges current support from DFG
grant Pu117/8-1 and earlier support from NASA ATP grant NNX11AC40G. VP
acknowledges support from fondes qu\'eb\'ecois de la recherche sur la
nature et les technologies. GAW acknowledges support from the Natural
Science and Engineering Research Council of Canada (NSERC).

\bibliographystyle{mn2e_fix}
\bibliography{sundqvist_macro}

\end{document}